\documentclass[useAMS,usenatbib]{mn2e}
\usepackage{amsmath, amssymb}
\usepackage{graphicx, times}
\pdfoutput=1
\usepackage{epstopdf}
\usepackage{float}

\title[Shock Breakout in a Wind]{Supernova Shock Breakout through a Wind}

\author[S. Balberg and A. Loeb]{Shmuel Balberg$^{1,2}$ and Abraham
Loeb$^1$\\ 
$^1$Harvard-Smithsonian Center for Astrophysics, 60 Garden
Street, Cambridge, MA 02138, \ USA\\
 $^2$Racah Institute of Physics, The Hebrew University, Jerusalem 91904, 
Israel}

\begin{document}
\maketitle

\begin{abstract}
The breakout of a supernova shock wave through the progenitor star's
outer envelope is expected to appear as an X-ray flash. However, if
the supernova explodes inside an optically-thick wind, the breakout
flash is delayed. We present a simple model for estimating the
conditions at shock breakout in a wind based on the general observable
quantities in the X-ray flash lightcurve: the total energy $E_X$, and
the diffusion time after the peak, $t_{\rm diff}$. We base the
derivation on the self-similar solution for the forward-reverse shock
structure expected for an ejecta plowing through a pre-existing wind
at large distances from the progenitor's surface. We find simple
quantitative relations for the shock radius and velocity at
breakout. By relating the ejecta density profile to the pre-explosion
structure of the progenitor, the model can also be extended to
constrain the combination of explosion energy and ejecta mass. For the
observed case of XRO08109/SN2008D, our model provides reasonable
constraints on the breakout radius, explosion energy, and ejecta mass,
and predicts a high shock velocity which naturally accounts for the
observed non-thermal spectrum.

\end{abstract}


\section{Introduction}
\label{sec:intro}

As a supernova shock wave propagates through the progenitor star, it
eventually emerges through the outer envelope region which has a low
optical depth. At this point the shock ``breaks out'' of the star,
initiating the first electromagnetic signal of the explosion. The
breakout results in a flash of radiation, as the internal energy
deposited by the shock diffuses out of the shocked region on a
timescale comparable to the dynamical time of the shock. Following
breakout the ejected envelope expands so that the photoshpere
recedes into the ejecta. Any remaining internal energy, which has not
been converted during adiabatic expansion, is then gradually radiated
to power the supernova lightcurve.

It has long been suspected that the shock breakout from a star would
appear as an X-ray flash \citep{Colgate74,KC78}, followed by a
UV/optical transient corresponding to emission from the outer layers
of the ejecta \citep{Falk78}. The interest in prompt signals from
supernovae increased recently due to the new capabilities of modern
searches for transients, such as
Pan-STARRS\footnote{http://pan-starrs.ifa.hawaii.edu} and
PTF\footnote{http://www.astro.caltech.edu/ptf/} as well as the planned
LSST\footnote{http://www.lsst.org/lsst} and
WFIRST\footnote{http://wfirst.gsfc.nasa.gov/}. In fact, several
supernovae lightcurves were observed early after the explosion both in
the UV and the optical bands
\citep{Gezarial08,Gezarial10,Soderbergal08}. In two cases, GRB060218/SN2006aj 
\citep{Campanaal06} and XRO08109/SN2008D \citep{Soderbergal08}, X-ray detectors  
on board the Swift satellite have captured a luminous X-ray outburst, which were 
later followed by observations of supernovae in the longer wavelengths. Both events 
may very well have been the signature of shock breakout.

These new observations motivated detailed theoretical modeling of the
early emission in supernovae. The models focused on the early
UV/optical lightcurve that follows breakout, on time scales of order a
day \citep{ChevFrans08,RW10,NS10}, for which the dynamics are somewhat
simpler (see also \citet{Chevalier92}).  Modeling the X-ray flash associated with
breakout is more complicated, since it must include the properties and
structure of the radiation-mediated shock (RMS) as it propagates
through the sharply decreasing density of the outer envelope, where
the shock width becomes non-negligible relative to the distance to the
edge of the star.  \citet{Katzal10} have shown that if the shock
velocity normalized by $c$, $\beta_S\geq 0.07$, free-free emission
will be unable to produce a large enough number density of photons to
establish a blackbody spectrum. As a result, the average photon energy
in the shocked region will be significantly larger than the
equilibrium blackbody photon temperature with the same energy
density. Hence, any analysis of the breakout lightcurve based on the
assumption of local thermodynamic equilibrium and a black body
spectrum will not reproduce the photon flux and spectrum emitted from
the shock region during breakout.
   
The shock breakout lightcurve could be significantly influenced by any
circumstellar material (CSM) surrounding the exploding star. In fact,
the aforementioned two candidates for an X-ray flash of a shock 
breakout origin are associated with a type Ibc supernova, in which the
progenitor is believed to be a Wolf-Rayet star that has undergone
significant mass loss prior to the explosion.  General, order of
magnitude estimates for the energy enclosed in shock breakout do
indeed favor a wind environment over a bare star in both cases
\citep{Waxmanal07,ChevFrans08,Katzal10}. If the supernova takes place
in an optically thick wind, the breakout is delayed from the
observer's point of view, and the X-ray flash emerges only when the
shock width in the wind becomes comparable to its distance from the
CSM photosphere. Even a dynamically unimportant wind (which does not
influence the emission of the early UV/optical lightcurve), will
completely change the properties of the X-ray flash relative to a bare
star by virtue of its optical depth.

Motivated by these considerations, we present here a simple analysis
that relates the observable properties of the X-ray flash to the
physical parameters of shock breakout in an optically thick
wind. While a self-consistent treatment of the shock structure, light
travel time and frequency dependence is required to reproduce the
lightcurve of the flash, the integral quantities should be
weakly dependent of these details. We apply the known self similar
solution of a supernova ejecta moving into a pre-existing wind
material with a density profile of $\rho\propto r^{-2}$
\citep{Chevalier82}, and demonstrate how it can be used to relate the
key observable features of the X-ray flash, the total energy $E_X$ and
the diffusion time scale $t_{\rm diff}$ to the parameters of the shock
breakout. Most notably, we can obtain a quantitative estimate of the
breakout radius and the associated shock velocity. Our model
expands upon the order of magnitude estimate presented by
\citet{Ofekal10} regarding PTF09uj, which possibly exhibited a long
($\sim 1$ week) UV flash due to breakout in a very dense wind. While
our model assumes a non relativistic shock, it can be generalized to
relativistic shocks and more complex wind structures, and also compared to
the specific model suggested by \citet{Li07} for SN2006aj (note,
however, that this source may have been significantly asymmetric,
\citep{Waxmanal07}).

The outline of the paper is as follows. In \S\ref{sec:shock} we review
the self similar solution for an ejecta expanding into a wind, and
discuss the applicability of this physical setting to the problem of
shock breakout. In \S\ref{sec:breakout_parameters} we present how the
self similar solution can be used to relate the observed properties of
the X-ray flash to the conditions at breakout.  In
\S\ref{sec:Explosion_parameters} we demonstrate that based on the
initial profile of the ejecta \citep{MM99}, the observable quantities
can be used to place constraints on the parameters of the underlying
explosion. In \S\ref{sec:SN2008D} we consider the particular example
of SN2008D, which is a borderline case in terms of the applicability
of our model.  Finally, \S\ref{sec:conclusions} summarizes our main
conclusions.

\section{The Forward-Reverse Shock Structure}
\label{sec:shock}

We consider a spherically symmetric supernova explosion in a nearly
stationary CSM, produced by a wind from the progenitor star. The wind
is assumed to be much slower than the ejecta. For a steady wind with a
mass loss rate $\dot{M}$ and a velocity $v_w$, the density profile
outside of the star is,
\begin{equation}\label{eq:rho_wind}
\rho(r)=\frac{\dot{M}}{4 \pi r^2 v_w}\equiv K r^{-2} .
\end{equation}

Following the explosion, the outgoing ejecta plows through the wind
material and slows down, driving a forward shock through the wind and
a reverse shock that propagates back into the ejecta. Once the shock
had propagated far enough from the star's surface so that the crossing
time and the size of the star can be neglected, one may assume an
asymptotic time-dependent density profile for the outer part of the
supernova ejecta of the form,
\begin{equation}\label{eq:rho_ejecta}
\rho_e=g^m r^{-m} t^{m-3} ,
\end{equation}
where $g$ and $m$ are constants which depend on the initial conditions
of the progenitor and the explosion. If both the ejecta and the wind
material are initially cold, a self-similar solution can be found to
describe for the structures of both the forward shock \citep{Parker63}
and the reverse shock \citep{Chevalier82}.  At any time, the flow
consists of a forward shock region of the swept up wind material, and
a reverse shock in a leading region of the ejecta. At the boundary
between the two regions, velocity and pressure are continuous but
density is not. Denoting the position of the forward shock by $R_F$,
the reverse shock by $R_R$ and the contact radius by $R_C$, the ratios
$R_R/R_F$ and $R_C/R_F$ are time-independent. The solutions are found
by integrating dimensionless functions of similarity variable,
$\eta=r^{1/\lambda}t^{-1}$, where $\lambda=(n-2)/(n-3)$. The forward
and reverse shock fronts serve as boundary conditions. Integration is
carried out up to $R_C$ from both ends, and the solutions are joined
by matching the velocities and pressures on both sides of the contact
surface.

\begin{figure}
\resizebox{\hsize}{!}
{\includegraphics{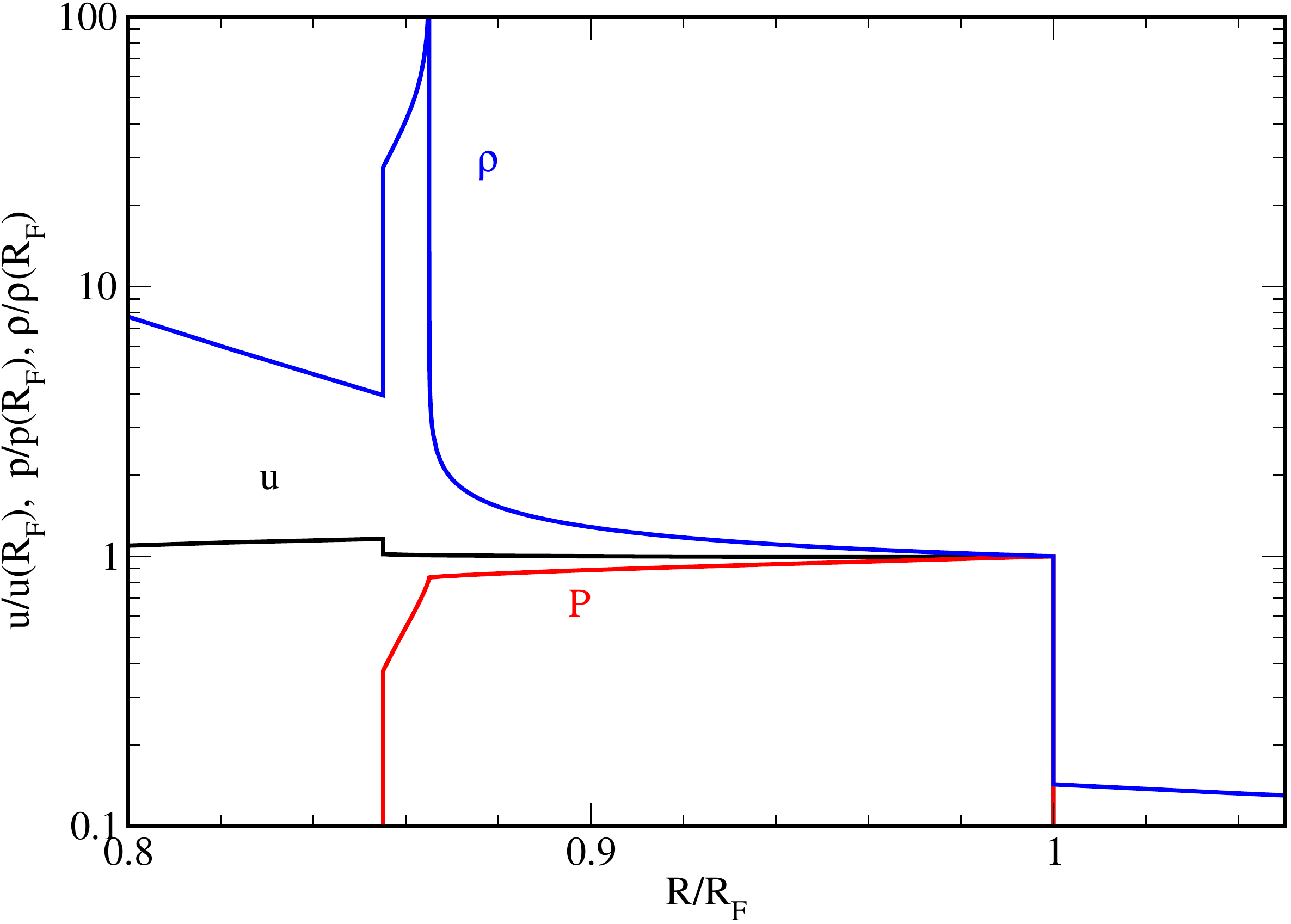}}
\caption[]{The velocity (black), pressure (red) and density (blue)
profiles of the self similar solution for supernova ejecta propagating
through a wind with $m=10$ and $\gamma=4/3$ (see text). The radial
coordinate is normalized to the radius of the forward shock, $R_F$, and
the physical quantities are normalized to their post shock values at the
forward shock front.
\label{fig:profile}}
\end{figure}

The self-similar solution describes the normalized profiles of the
density, velocity and pressure profiles as a function of $\eta$, and
depends on the values of $m$ and the adiabatic index of the shocked
materials, $\gamma$. An example of such a profile for $m=10$ and
$\gamma=4/3$ is presented in Figure \ref{fig:profile}, where all
quantities are normalized to their values just behind the forward
shock. The gradual divergence of density at the discontinuity (in both
of the forward and reverse shock regions) is generic to the assumed
$r^{-2}$ wind profile \citep{Chevalier82}, for which the solutions
lead to a vanishing speed of sound on both sides of the contact
radius.  In reality, the divergence will be avoided since the density
discontinuity is Rayleigh-Taylor unstable and hence the contact region
will obtain a finite width.

Our choice of $\gamma=4/3$ is, of course, motivated by our assumption
that the shocks involved are radiation dominated; the internal energy
density in the shocked region is therefore $\varepsilon=3P$, where $P$
is the pressure (see \citep{Chevalier83} for comparison, with $\gamma=4/3$ 
motivated by cosmic-ray dominated shocks). 
By integrating over the pressure in the solution we
can find the total energy in radiation. The pressure behind the
forward shock, $P_F$, is determined by the strong shock condition,
\begin{equation}
\label{eq:P_F}
P_F=\frac{6}{7}K R^{-2}_F \dot{R}^2_F ,
\end{equation}
where $\dot{R}_F$ is the velocity of the forward shock. Denoting the
relative width of the shocked region by $x=(R_F-R_R)/R_F$ we may
express the total internal energy in the shocked region as:
\begin{equation}
\label{eq:E_rad}
E_{rad}(x,R_F)=\int_{R_R}^{R_F}4\pi r^2 \varepsilon (r) dr= \chi_E \frac{72\pi}{7}K R_F x \dot{R}^2_F .
\end{equation}
The dimensionless quantity $\chi_E$ is defined by
\begin{equation}\label{eq:Chi_E}
\chi_E=\frac{\int_{R_R}^{R_F}4\pi r^2 P (r) dr}{4 \pi R^3_F x P_F} ,
\end{equation}
and is determined by the details of the post-shock profile.  For the
particular case of $m=10$ and $\gamma=4/3$ depicted in Figure
\ref{fig:profile}, $x=0.146$, $\chi_E=0.9$.

The total width $x$ is dominated by the forward shock region, which
has a width of $x_F=(R_F-R_C)/R_F=0.135$, while the width of the
reverse shock region is only $x_R=(R_C-R_R)/R_F=0.011$. However, the
density in the reverse shock is much larger, and we find that
$M_R/M_F=2.23$, where $M_R$ and $M_F$ are the masses enclosed in the
reverse and forward shock regions, respectively. By construction,
$M_F=4 \pi K R_F$. The resulting $(M_R+M_F)\approx 3.2M_F$ implies
that order of magnitude estimates identifying the total energy of the
flash with $M_F \dot{R}^2_F/2$ (see, e.g., \citet{Katzal10}) should be
corrected. While it is clear that the mass of the ejecta affected by
the wind must be comparable to that of the swept-up wind, ignoring the
mass enclosed in the reverse shock leads to a significant overestimate
of the shock velocity.

The density profile in the self similar solution can be integrated to find
the normalized column density in the forward and reverse shock
regions, which can then be used to estimate the optical depth and
the diffusion time scales.  First, we define,
\begin{equation}
\label{eq:eff_tau}
\hat{\tau}_F=\frac{\int_{R_C}^{R_F} \rho(r) dr }{7 K R^{-2}_F(R_F-R_C)}\;,\; ; 
\hat{\tau}_R=\frac{\int_{R_R}^{R_C} \rho(r) dr }{7 K R^{-2}_F(R_C-R_R)} .
\end{equation}
The results for the self similar solution depicted in Figure
\ref{fig:profile} are $\hat{\tau}_F\simeq 1.3$ and $\hat{\tau}_R \simeq 4.2$.  The
total optical depth of the shock region is therefore
\begin{equation}
\label{eq:tau_S}
\tau_S=\kappa\left(\hat{\tau}_F x_F R_F+\hat{\tau}_R x_R R_F\right)7 K R^{-2}_F \simeq 1.55 \kappa K R^{-1}_F ,
\end{equation}
where we have assumed a uniform, frequency independent opacity,
$\kappa$. The total diffusion time scale in the shocked region,
$t_{DS}$ is then,
\begin{equation}
\label{eq:t_diff_shock}
t_{DS}=\tau_S \frac{x R_F}{c}\simeq 0.23 \frac{\kappa K}{c} .
\end{equation}

Finally, the physical scales of the full solution are determined by
the dimensional parameters, $g$ and $K$. In particular, the velocity
of the forward shock is related to its position by,
\begin{equation}
\label{eq:R_dot of R}
\dot{R}_F=\frac{m-3}{m-2}\left(\frac{A_F g^m}{K}\right)^{1/(m-3)}R_F^{-1/(m-3)} ,
\end{equation}
where $A_F$ is another constant that is extractable from the
self-similar solution \citep{Chevalier82}; for $m=10$ and $\gamma=4/3$
we get $A_F\approx0.07$.

We note that the numerical values of $x$, $\chi_E$, $\hat{\tau}_F$ and
$\hat{\tau}_R$ are very weakly dependent on $m$. In the range of
$m=8$--$12$, we find that these values vary by no more than $\pm 20\%$
with respect to their values for $m=10$, which we use for reference
below. The value of $A_F$ is not as well constrained, and changes
within a factor of 2 in this range. However, as shown in
\S~\ref{sec:Explosion_parameters}, the final result is not sensitive
to this uncertainty.

\section{A Simple Parametrization of the Breakout Conditions}
\label{sec:breakout_parameters}

We now use the results of the previous section to estimate the shock
breakout features.  We assume that prior to shock breakout the flow is
approximately adiabatic in the shocked regions, and that leakage of
radiation into the cold parts of the wind and ejecta is negligible. We
discuss the validity of these assumptions below.

We take the total energy observed in the breakout flash, $E_X$, to be
comparable with the energy carried in radiation in the shocked region,
$E_{rad}$. Similarly the time scale characterizing the post maximum
part of the lightcurve is determined by diffusion from the shock front
through the unshocked wind, $t_{\rm diff}$. These associations are
valid because the diffusion time of radiation in the shocked
region is shorter than the time scale for diffusion from the shock to
the unshocked wind's photosphere. The
optical depth from the shock front in the wind material is
$\tau=\kappa K/R_F$, and so the time scale for diffusion from the
shock front is
\begin{equation}\label{eq:t_diff}
t_{\rm diff}\approx \tau \frac{R_F}{c}=\frac{\kappa K}{c} ,
\end{equation}
or about four times longer than the timescale for diffusion in the
shocked region, given by equation (\ref{eq:t_diff_shock}). The opposite
regime of $t_{\rm diff} \ll t_{DS}$ would have been characterized by a
longer, fainter lightcurve.

Observationally determined values of $E_X$ and $t_{\rm diff}$ and the self
similar solutions provide the necessary relations for estimating the
key properties of ejecta-wind configuration at breakout. First, using
Equation (\ref{eq:t_diff}) the observed value of $t_{\rm diff}$ is
sufficient to determine the value of the wind density constant, $K$:
\begin{equation}
\label{eq:K}  
K= \frac{c t_{\rm diff}}{\kappa}=1.5\times 10^{14} 
\frac{t_{{\rm diff},3}}{\kappa_{0.2}}~{\rm g~cm^{-1}} ,
\end{equation}
where $t_{{\rm diff},3}=(t_{\rm diff}/10^3~{\rm s})$ and
$\kappa_{0.2}=(\kappa/0.2~{\rm cm^2~g^{-1}})$ (the fiducial opacity
normalization is appropriate for wind material that is dominated by
helium and heavier elements.)  The value of $K$ for $t_{{\rm
diff},3}=1$ corresponds to a wind mass loss rate of about $3\times
10^{-4} M_{\sun}~{\rm yr}^{-1}$ for a wind velocity of
100~km~s$^{-1}$.

The criterion for shock breakout is that the diffusion time be equal
to dynamical time, $t_{\rm diff}=t_{\rm dyn}\equiv R_F/\dot{R}_F$. We use this 
relation to substitute the shock
velocity from Equation (\ref{eq:E_rad}) for $t_{\rm diff}$, yielding
\begin{equation}
\label{eq:E_radnew}  
E_X= \frac{72 \pi c }{7 \kappa} \chi_E x R^3_S t^{-1}_{\rm diff} .
\end{equation}
Using the predetermined values of $\chi_E$ and $x$ based on the
self-similar solutions in \S \ref{sec:shock}, the breakout radius can be 
expressed in terms of $E_X$ and $t_{\rm diff}$:
\begin{multline}\label{eq:R_BO}
R_{BO}=\\
5.4\times 10^{12} \left(\frac{E_{X,47}t_{{\rm diff},3}}{\kappa_{0.2}}\right)^{1/3}
\left(\frac{x}{0.146}\right)^{-1/3}\left(\frac{\chi_E}{0.9}\right)^{-1/3} {\rm cm} ,
\end{multline}
where $E_{X,47}=(E_X/10^{47}~{\rm ergs})$. We emphasize again that the
specific values of $x$ and $\chi_E$ found in the self-similar
solutions are weakly dependent on the exact value of the ejecta
density profile, $m$. Combined with the fact that the breakout
radius depends only on the cubic roots of these numerical factors, we
conclude that the the result of Equation (\ref{eq:R_BO}) is
robust to uncertainties in the details of the ejecta profile.

Using the inferred values of the breakout radius and the wind density 
coefficient, $K$, we can also express the masses of the wind material 
and the ejecta included in the forward and reverse shock, respectively.
The mass in the forward shock is:
\begin{multline}\label{eq:M_BO}
\frac{M_F}{M_\odot}= \\
5.1\times 10^{-6} E^{1/3}_{X,47}t^{4/3}_{{\rm diff},3}\kappa_{0.2}^{-4/3}
\left(\frac{x}{0.146}\right)^{-1/3}\left(\frac{\chi_E}{0.9}\right)^{-1/3},
\end{multline}
and $M_R$ is about $2.23$ times larger. It is a useful consistency check that  
the mass of the ejecta involved in the shock formation is significantly smaller than 
the total ejecta mass. An ejecta density profile of the form of equation (\ref{eq:rho_ejecta}) 
is only applicable to the edge of the progenitor envelope, where the initial density 
drops sharply with radius. 

An important consequence of the estimate for the breakout radius is
the corresponding shock velocity. The value of this
velocity (normalized by $c$) is, $\beta_s \sim
0.18E^{1/3}_{rad,47}t^{-2/3}_{{\rm diff},3}$. This relates the
observed global features of the lightcurve and the spectrum, since for
$\beta>0.07$ the photon generation rate in the shocked region is
insufficient to establish a Planck distribution within the shock width
\citep{Katzal10}. The emergent spectrum in that regime is dominated by
harder ($\geq 1$keV) X-ray photons than a blackbody, since the same
amount of energy is shared by fewer photons. If, on the other hand,
$\beta_s < 0.07$, the flash should exhibit a blackbody spectrum
with a temperature of
\begin{multline}\label{eq:T_BB}
T_{BB}\approx \\
2.64\times 10^5 E^{1/12}_{rad,47} 
t^{-5/12}_{{\rm diff},3} \kappa^{1/6}_{0.2} 
\left(\frac{x}{0.146}\right)^{1/6}\left(\frac{\chi_E}{0.9}\right)^{1/6} 
{\rm K},
\end{multline}
assuming that the characteristic luminosity is $L=4\pi R^2_{BO} \sigma
T^4_{BB}\sim E_X/t_{\rm diff}$.

Some cautionary remarks should be mentioned about our results. First,
a useful aspect of equation (\ref{eq:R_BO}) is that for the reference
values of $E_{rad}\sim 10^{47}$ergs and $t_{\rm diff} \sim 10^3$s, the
light crossing time at the breakout radius is about $18\%$ of the
diffusion time. Therefore, the geometric time delay due to the light
crossing time will have a secondary, but not negligible, effect on the
earlier part of the flash lightcurve. Another caveat is related to our
assumption that the kinetic energy of the shocked region (and that of
the unshocked ejecta) will not be converted into internal energy over
the time scale of the breakout flash. We expect this approximation to
be valid, since the shocked region would slow down significantly only
after it sweeps up an additional wind mass comparable to its own,
$(M_R+M_F)=3.2\times 4 \pi K R_{BO}$ (see \S \ref{sec:shock}). Hence,
it takes at least several dynamical times, or diffusion times, to
convert kinetic energy to internal energy. Moreover, the kinetic energy of the
shocked material, $E_{kin}\approx\frac{1}{2}(M_R+M_F)(\dot{R}_F/7)^2$, is only 
about 10\% of its internal energy, and so the addition to the available 
energy would generate at most a 3\% correction in the value of $R_{BO}$ in equation 
(\ref{eq:R_BO}). The main reservoir of kinetic energy lies in the yet unshocked 
ejecta, but this region takes even longer to slow down and is initially hidden 
behind larger optical depths.

Another secondary effect we have neglected is the possible
acceleration of gas ahead of the shock by radiation which escapes the
shocked region.  While an exact calculation of the extent of this
effect requires a radiation-hydrodynamics simulation, it is
straightforward to demonstrate that relative to velocities 
$\gtrsim 10^9 {\rm cm\;s}^{-1}$ significant acceleration of wind material 
at any given radius below $R_{BO}$ cannot take place before the 
shock overtakes that radius. It is also noteworthy 
that the optical depth of the unshocked ejecta is quite large, so radiative 
effects should not increase the width of the reverse shock significantly 
prior to breakout.

A more serious concern is to what extent the ejecta is in a 
cold, power law density density state prior to breakout. 
For compact progenitors with a radius $R_\star \sim 10^{11}$cm, 
the reference values give $R_{BO}/R_\star\approx 54$. While this ratio 
is sufficiently large to justify neglecting the progenitor radius as a 
relevant scale in the solution, it is not obvious that the density profile 
should already settle at $R_{BO}$ into the asymptotic form we assumed in equation
(\ref{eq:rho_ejecta}). Originally, the supernova shock wave deposits
in the outer envelope $\sim 6$ times as much internal energy as
kinetic energy. For spherical adiabatic expansion, the internal energy
declines as $E\sim r^{-1}$, so for $(E_{rad,47}t_{{\rm diff},3})<1$ the ejecta can 
still be ``lukewarm'' rather than cold, implying that the ejecta density profile may 
still be developing, and perhaps that the reverse shock may not be very strong. 
This might lead to some deviation of the gas dynamics from our simple model.

\section{A Simple Parametrization of the Explosion Properties}
\label{sec:Explosion_parameters}

We next incorporate the properties of the explosion
into our model. This can be done by relating the two quantities
describing the ejecta profile, $g$ and $m$, to the explosion
parameters. In practice, obtaining such a relation requires a
specific model for the fast material leading the ejecta, and we use
the standard model of \citet{MM99}.  In this model, the original
density in pre-explosion outer layers of the star, $\rho_0$, is a
simple function of the relative distance from the edge of the star,
\begin{equation}\label{eq:rho_delta}
\rho_0(r_0)=\rho_\star\delta^n\;\; ; \delta=1-r_0/R_\star ,
\end{equation}
where $R_\star$ is the outer radius of the progenitor and and $r_0$ is the
radial position within the progenitor. The power $n$ is a constant
which depends on the assumed physics of the envelope -- $n=3$ for a
convective envelope and $n=3/2$ for a radiative one, and the density
scale, $\rho_\star$, depends on the details of the progenitor structure.

Given this initial density profile, \cite{MM99} have shown that after 
the passage of the shock the velocity of the material in the 
very outer envelope follows an approximate dependence on the 
energy of the explosion, $E$, and the total mass of the ejecta, $M$:
\begin{equation}\label{eq:v_delta}
v(r_0)=A_v \left(\frac{E}{M}\right)^{1/2}\left(\frac{4 \pi}{3
f_\rho}\right)^\beta \delta^{-\beta n} ,
\end{equation}
where $\beta\approx 0.19$ is practically independent of the details of
the progenitor and the explosion, and $f_\rho\equiv
\rho_\star/\bar{\rho}$, where $\bar{\rho}=3M/4\pi R^3_{\star}$
is the average mass density within the star. The asymptotic value of
the coefficient $A_v$ is $\sim 2 A_{v,S}$, where $A_{v,S}\approx
0.79$ is the appropriate value for the {\it shock} velocity profile at
the onset of the explosion. If the wind were dynamically unimportant
and the ejecta could reach reach this asymptotic velocity profile,
then the resulting density profile obtains the form \citep{RW10},
\begin{equation}\label{eq:rho_ejecta_delta}
\rho(\delta_m,r)=-\frac{M}{4\pi r^3}\frac{(n+1)}{\beta n} \delta_m ,
\end{equation}
where $\delta_m(\delta)$ is the Lagrangian mass fraction of an element
of the ejecta which was initially at $\delta$ in the pre-explosion
profile. With $\delta_m=3 f_\rho \delta^{(n+1)}/(n+1)$ and approximating that
an element with $\delta_m$ reaches a radius $r$ at a time of $t\simeq
r/v$ we arrive at an expression of the ejecta in the form of equation
(\ref{eq:rho_ejecta}):
\begin{equation}\label{eq:rho_ejecta_fin}
\rho(r,t)=\left(\frac{4\pi}{3
f_\rho}\right)^{1/n}(A_v)^{2\theta}E^\theta M^{1-\theta}r^{-m}t^{m-3} ,
\end{equation}
where $\theta=(n+1)/(2\beta n)$, and 
\begin{equation}\label{eq:m=}
m=-3-\frac{n+1}{\beta n} .
\end{equation}
Again, we assume that the ejecta is cold and has developed the
density profile of equation (\ref{eq:rho_ejecta}).  Joining equations
(\ref{eq:rho_ejecta_fin}) and (\ref{eq:R_dot of R}) sets a
quantitative constraint on the combination of the explosion energy,
$E$, and the ejecta mass, $M$. Specifically for $\beta=0.19$ and
$n=3$ we have $m\simeq 10$ and $\theta\simeq 3.5$, and with the aid of
the characteristic values derived in the self-similar solution we
arrive (after some algebra) at the final result:
\begin{multline}\label{eq:EM_const}
\left(\frac{E}{10^{51}~{\rm
    ergs}}\right)\left(\frac{M}{M_\odot}\right)^{-5/7}\simeq \\
0.8 A^{-2}_v f^{2/21}_\rho (\frac{A_F}{0.07})^{-2/7} \kappa^{10/21}_{0.2}
E^{16/21}_{rad,47}t^{-20/21}_{{\rm diff},3} .
\end{multline}
Although this equation is approximate, it implies that within the
framework of our model, the plausible range of values for the entire
prefactor in equation (\ref{eq:EM_const}) is rather limited. The
values of $x$ and $\chi_E$ are tightly constrained in the self-similar
solution, and do not introduce a significant uncertainty. The
coefficient $A_F$ which does depend on the power of the ejecta density
profile is suppressed by the 2/7 power, and the unknown factor
$f_\rho$ is suppressed by the 2/21 power. Finally, $A_v\approx 1.6$
for ejecta that has approached its asymptotic velocity. The resulting
equation can be applied as a sanity check for the value of the product
$E M^{-5/7}$, when a flash is suspected to be the result of a
supernova-wind breakout scenario.

\section{The X-ray Flash in XRO080109/SN2008D}
\label{sec:SN2008D}

We now examine the predictions of our model for the best candidate to date 
of an X-ray flash due to shock breakout in a wind XRO080109, associated 
with the type Ibc supernova SN2008D \citep{Soderbergal08}. 
Whereas a similar analysis can be applied to the case of GRB060218/SN2006aj,
this source requires relativistic expansion \citep{Li07}, and possibly 
a significant deviation from spherical symmetry \citep{Waxmanal07}, so we 
do not consider it here.

The bright X-ray transient XRO080109 was serendipitously discovered
during a scheduled Swift Telescope observation of the galaxy NGC2770
(at a distance of $\sim 27$ Mpc). The observed spectrum shows a
non-thermal shape with a power-law frequency dependence of the photon
number flux per unit frequency, $N_\nu\propto \nu^{-\Gamma}$ with
$\Gamma=2.3\pm 0.3$, through the observed range of X-ray energies,
0.3--10keV. The follow-up observation by the Ultraviolet/Optical
Telescope (UVOT) on board Swift discovered a counterpart Type Ib
supernova, denoted SN2008D. Analyses of the main lightcurve of SN2008D
\citep{Mazzalial08,Modjazal09,Tanakaal09} favor an underlying
explosion of a compact progenitor, presumably a Wolf-Rayet star due to
the lack of hydrogen lines in the lightcurve. A consistent result was
found by \citet{RW10} in their analysis of the early UV/optical
lightcurve, in which they inferred a initial progenitor radius of
$\sim 10^{11}$cm. Note that for such a compact progenitor, a typical wind 
velocity should be of order $1000~$km s$^{-1}$ 
(similar to the escape velocity).

The lightcurve of XRO080109 displayed a rapid rise and an exponential
decline, with an $e$-folding time in the declining phase of
approximately $t_e=130$s.  Such a timescale is too long to be consistent 
with breakout through the surface of a bare star, and the most plausible 
explanation links this decline with the diffusion of the radiation from the shock 
through the unshocked wind, $t_e=t_{\rm diff}$, which we will use below. 
For an alternative interpretation, that the lightcurve was shaped by an
aspherical explosion, see \citet{Couchal10}.

Quantitative estimates of the total energy of the burst (assuming
isotropic emission) depend on the modeling of the column density near
the source and extinction; \citet{Soderbergal08} estimated the total
energy in the burst to be $E_{X}\approx 2\times
10^{46}$ergs, for which the appropriate breakout radius arising from
equation (\ref{eq:R_BO}) is $R_{BO}\approx 1.6\times 10^{12}$cm. For
the lower value of $E_{X} \approx 6\times 10^{45}$ergs inferred by
\citet{Modjazal09}, the breakout radius is reduced to $R_{BO}\approx
1.1\times 10^{12}$cm. In either case, the breakout must have occurred
in a wind rather than on the surface of a bare Wolf-Rayet progenitor.
Our result is consistent with the lower limit of $7\times 10^{11}$cm for 
the breakout radius found by \citet{Soderbergal08}, based on the fact that 
the thermal component of the X-ray flash lies below the XRT low energy 
cutoff of $\sim0.1\;$keV. 

We note that the diffusion time scale indicates a density profile with
$K\simeq 1.95\times 10^{13}$g cm$^{-1}$, corresponding to
$\dot{M}/v_w\approx 3.85 \times 10^{-4} M_{\sun} {\rm yr}^{-1}/(1000
{\rm km~s^{-1}})$. This value is an order of magnitude larger than 
the range estimated by \cite{Soderbergal08} and \citet{ChevFrans08} from 
the radio observations of SN2008D, assuming a 10\% efficiency in magnetic 
field production behind the shock front. Taken at face value, our result 
implies a lower efficiency for magnetic field production, or an enhanced mass 
loss rate just prior to explosion, or both. We note that that \citet{Soderbergal08}
estimated a low $\dot{M}$ also from the breakout flash, but this discrepancy 
is alleviated once our value for the breakout radius is used (rather than their 
lower limit), along with with a specific opacity of $\kappa=0.2\;$cm$^2$g$^{-1}$ 
and a remaining optical depth of $\tau=$2--3 in the unshocked wind.

The corresponding total amount of mass involved in the forward and reverse 
shocks at breakout comes out to be of order $10^{-7}M_\odot$. Although small, 
this mass is several orders of magnitude larger than that enclosed in the 
last few optical depths of a bare Wolf-Rayet star with 
$R_\star\approx 10^{11}$cm 
\citep{MM99}, again supporting the wind breakout scenario.

Finally, the combination of explosion energy and ejected mass in
equation (\ref{eq:EM_const}) yields $(E/10^{51}{\rm
ergs})(M/M_{\sun})^{-5/7}\approx 0.26--0.64 f^{2/21}_\rho$ for the total 
energies of $E_X=0.6$--$2\times 10^{46}$ergs. Depending on the value of 
$f_\rho$, this result appears to lie between the combination originally 
suggested by \citet{Soderbergal08} of 
$(E/10^{51}{\rm ergs})=2,\;(M/M_{\sun})=5$, and the more energetic explosions
inferred later by \citet{Mazzalial08} and
\cite{Tanakaal09}, $(E/10^{51}{\rm ergs})=6,\;(M/M_{\sun})=6$--$8$.

We caution that $(R_{BO}/R_\star) \sim 11$--$16 (R_\star /10^{11}{\rm
cm})^{-1}$ is modest, and so the deviation from the cold ejecta model
may be significant (and even more so if the progenitor radius was a
few $10^{11}$cm). This deviation does not change much the
energetics of the model, which is why the general results are
consistent with shock breakout in a wind, but the details of the shock 
structure are most likely different than those predicted by our solution. 
A further complication will arise from the wind density profile, which close 
to the star may be significantly steeper than a $\rho\sim r^{-2}$ dependence 
\citep{Li07}. If the shocked wind material is dominated by this component, 
additional quantitative corrections must be included in our model. Such a wind 
structure may also provide a natural explanation the relatively large mass loss 
rate we required assuming $\rho\sim r^{-2}$ throughout the entire wind.

Another point of interest is that the shock velocity at breakout we find for 
SN2008D is quite large, with $\beta\approx 0.3$--$0.4$.  This range is still 
consistent with our non-relativistic approach, and is also in agreement with 
the velocity range inferred from later VLBI measurements 
\citep{Bietenholzal09}. However, given the relatively small 
breakout radius, it is most likely an upper limit:  
at $(R_{BO}/R_\star)\approx 10$--$15$  the ejecta 
may still be carrying a significant fraction of its initial internal energy 
that was not converted to kinetic energy. 
We also emphasize that if the optical depth at breakout is only
$\tau=\beta^{-1}\approx 2.5-3$, the diffusion approximation becomes
questionable. Finally, at $\beta>0.2$ pair creation in the shock is
non-negligible \citep{Weaver76,Katzal10,Budnikal10}, causing the
equation of state to deviate from a pure $\gamma=4/3$ ideal
gas. Nonetheless, our simple model is robust enough to demonstrate
that the shock velocity in SN2008D was large enough for the photon
energy distribution not to be Planckian, leading naturally to the
observed non-thermal X-ray spectrum.

\section{Conclusions and Discussion}
\label{sec:conclusions}   

We have presented a simple model for the properties of shock breakout
in a supernova explosion embedded within an optically thick wind. By
assuming that the ejecta and wind form a forward-reverse shock
structure \citep{Chevalier82}, we have shown that the breakout radius
can be estimated based on the observationally determinable values of
the total energy in the flash and the diffusion time scale from the
shock to the wind's photosphere. Among other things, we demonstrated
that the time scale for photon diffusion in the shocked region is
shorter than the time scale for photon diffusion from the shock to the
photosphere, justifying the underlying assumption that breakout will
manifest itself as a brief flash even in the presence of an optically
thick wind.  We also found that if the ejecta time-dependent density
profile is derived from the initial progenitor model of \citet{MM99},
the combination of explosion energy and ejecta mass can be estimated
from the observed quantities of the shock breakout. For a low flash
energy and long diffusion time, a blackbody spectrum is expected to
emerge. In the opposite regime, free-free emission cannot
generate a sufficient number photons to establish a thermal spectrum 
during the shock passage \citep{Katzal10}.

While limited by its simplifying assumptions, our model describes the
general features of a shock breakout in a wind around an exploding
Wolf-Rayet star, characterizing a Type Ibc supernova. We applied our model
to the X-ray flash observed in association with SN2008D \citep{Soderbergal08}
and found sensible values for the shock breakout radius ($\sim 1.1$--$1.6\times
10^{12}$cm) as well as the properties of the entire explosion. 
However, such a breakout radius is probably not sufficiently larger than 
the progenitor radius for our model to be fully applicable. It is therefore 
possible that we are overestimating the shock velocity at breakout, which we 
found to be $\beta_s\sim0.3$--$0.4$. Nonetheless, our
model clearly demonstrates that if XRO080109 was indeed shock breakout
in SN2008D, the shock velocity was certainly large enough to prevent a
blackbody spectrum from developing, in agreement with observations.

{\it Additional note:} After the completion of this paper a related preprint 
was posted by \citet{ChevalierIrwin11}. They too use the self-similar 
solution to assess the total energy available for shock breakout 
through a wind, but their work focuses on a different setting 
involving a red supergiant exploding into a much denser wind.
In their case the typical breakout radius is $R_{BO}\sim 10^{15}$cm and 
the mass of the ejecta involved in the shock comes out to be $>1M_\odot$; 
hence the assumption of a simple power law dependence for the density of the 
ejecta (equation \ref{eq:rho_ejecta}) may not be adequate 
(see \S~\ref{sec:breakout_parameters}).
 
\section*{acknowledgments}
We wish to thank Alicia Soderberg for valuable comments and Udi Nakar and Eli 
Waxman for helpful discussions. We also thank Roger Chevalier for
his helpful comments as the referee of this paper. SB thanks the Institute for 
Theory \& Computation (ITC) at Harvard
University for its kind hospitality.  This work was supported in part
by NSF grant AST-0907890 and NASA grants NNX08AL43G and NNA09DB30A.

\bibliography{refs_BOwindmod2}

\end{document}